\documentclass{mhd}     
\usepackage{graphicx}       

\usepackage{amssymb}
\usepackage{amsmath}
\usepackage{graphicx}       

\newcommand\Rm{{\rm Rm}}

\newcommand\Ru{{\rm Re}}

\newcommand{\bu}{{\bf u}}  

\newcommand{\bx}{{\bf x}}
\newcommand{\bZ}{{\bf z}}
\newcommand{\bb}{{\bf b}}  
\newcommand{\bJ}{{\bf J}}  

\newcommand{\bF}{{\bf F}}  

\newcommand{\bk}{{\bf k}}

\newcommand{\be}{{\bf e}}

\title{Cross helicity sign reversals in the dissipative scales of magnetohydrodynamic turbulence}

\author{V.~Titov\inst{1},  R.~Stepanov\inst{1},  N.~Yokoi\inst{2}, M.~Verma\inst{3}, R.~Samtaney\inst{4}}

\institute{Institute of Continuous Media Mechanics, Perm, 614013, Russia  \and Institute of Industrial Science University of Tokyo, Tokyo 153-8505, Japan \and  Department of Physics, Indian Institute of Technology Kanpur, Kanpur 20816, India \and Mechanical Engineering, Division of Physical Science and Engineering, King Abdullah
University of Science and Technology, Thuwal 23955-6900, Kingdom of Saudi Arabia }

\begin{document}

\maketitle
\begin{abstract}%
We perform direct numerical simulations of magnetohydrodynamic (MHD) turbulence with kinetic energy and cross helicity injections at large scales.  We observe that the cross helicity changes sign as we go from large and intermediate scales to small scales. In addition, the magnetic reconnections are strongest at the regions where the cross helicity changes sign and becomes smallest in magnitude. Thus, our simulations provide an important window to explore the regions of magnetic reconnections in nonlinear MHD.
\end{abstract}


\section{Introduction.}

The concept of magnetohydrodynamic (MHD) turbulence is important to explain
a great number of plasma phenomena observed in fundamental astrophysics and applied physics fusion plasma. Studies of MHD turbulence are aimed to achieve the common points of view on excitation, nonlinear transport and dissipative mechanisms. Still there are many aspects that need to be clarified. Magnetic reconnection is a crucial physical phenomenon that significantly affects the energy release during many astrophysical processes, such as solar flares, aurora in the Earth's magnetosphere, etc. This is the topic of the present paper.

A simple MHD model for a steady state of magnetic reconnection was suggested by Parker \cite{SP1957}. In the presence of turbulence, the reconnection rate may be drastically changed as compared with the laminar case. The primary effect of turbulence is to enhance the effective transport. At large magnetic Reynolds numbers, which are ubiquitous in astrophysical phenomena, the turbulent magnetic diffusivity in the magnetic induction is much larger than the molecular counterpart such as the Spitzer diffusivity. Through this effective diffusivity turbulence is expected to contribute to the enhancement of magnetic reconnection.

However, in the presence of a broken {mirror symmetry caused by a kind of external helical forcing}, we have other important turbulence effects. In addition to the enhanced transport, suppression of transport may be also caused by the turbulence. The total amount of the cross helicity (velocity--magnetic-field correlation), which represents the asymmetry between the directions parallel and antiparallel to the magnetic field, is an inviscid invariant of the MHD equations as well as the counterparts of the MHD energy and the magnetic helicity. The basic role of the cross helicity in the magnetic induction is to suppress the turbulent magnetic diffusivity arising from the turbulent energy \cite{yokoi2013a}. Recently it is found that a weak inverse transfer of kinetic energy toward the large scales can be due to a presence of strong cross helicity at small scales \cite{2019arXiv190105875B}.


It was found that in the presence of the turbulent cross helicity, the dynamic balance between the transport enhancement and suppression occurs~\cite{Yokoi2013}. These turbulence effects on transport are based on the coarse-grained view or the mean-field approach to turbulence. However, this does not mean that the dynamic-balance arguments are not relevant to full direct numerical simulations (DNSs) of the fundamental MHD equations. These basic properties of turbulence transport are intrinsically contained in the fundamental equations of the MHD turbulence with broken symmetry. The applicability of the mean-field approach in the context of magnetic reconnection was shown by numerical simulations of a Reynolds-averaged MHD turbulence model \cite{PhysRevLett.110.255001}. A subgrid-scale (SGS) turbulence model was used to investigate the contributions of the SGS turbulent energy and cross-helicity to the plasmoid reconnection rate \cite{Widmer2016}.

If the cross-helicity density changes its sign at some spatiotemporal region at some scale, there is no transport suppression in this region because of the vanishing cross helicity. Rather we have a transport enhancement originating from the turbulent energy. This lack of the transport suppression readily leads to a maximised magnetic reconnection due to a localised enhanced turbulent diffusivity \cite{Yokoi2013}. Once magnetic reconnection is induced at some point with the associated local out- and in-flows, a quadrupole spatial distribution of the cross helicity is reproduced and sustained. Then a quadruple cross-helicity configuration with the reversal of the cross-helicity sign at the symmetric points, preferable for the enhanced magnetic reconnection, is likely to be further reinforced \cite{yokoihoshino2011}. This suggests strong correlations between the regions  of  cross-helicity reversal and those of  strong reconnections with non-zero cross helicity.

The turbulent magnetic diffusivity or resistivity in the correlated (non-zero cross helicity) MHD turbulence has been investigated for a long time. Using numerical simulations of the eddy-damped quasi-normal Markovianised (EDQNM) approximation closure equation, it was shown that the sign of the cross helicity reverses at the dissipation scale with the equipartition between the kinetic and magnetic energies \cite{grappin1983}. With the aid of direct numerical simulations of the decaying two-dimensional MHD flows, the spectral properties of MHD turbulence with a non-zero cross helicity were investigated and the current sheet formation at the very small scales was reported \cite{politano1989}. Field theory calculation for MHD turbulence with large cross helicity was performed in \cite{Verma:PR2004}.

The purpose of this paper is to demonstrate that the scenario for reconnection of magnetic field is realised under conditions of homogeneous isotropic forced MHD turbulence. We found that, starting with a small scale in the dissipative interval, an isolated reconnection structures are observed which are well correlated with regions where cross helicity changes its sign.

\section{Direct numerical simulation of MHD turbulence with cross helicity injection.}

We solve the dimensionless equations of incompressible magnetohydrodynamics
\begin{eqnarray}
\partial_t{\mathbf{u}}  + ({\bf u} \cdot \nabla) {\bf u} & = &    ( {\bf b} \cdot \nabla) {\bf b} + \Ru^{-1}\,\, \nabla^2 {\bf u} + {\bf F}_u -  \nabla p, \label{eq:MHD_formalism:MHDu} \\
\partial_t{\mathbf{b}} + ({\bf u} \cdot \nabla) {\bf b}   & = &  ({\bf b} \cdot \nabla) {\bf u}   + \Rm^{-1} \nabla^2 {\bf b} + {\bf F}_b,  \label{eq:MHD_formalism:MHDB} \\
\nabla \cdot {\bf u} = \nabla \cdot {\bf b} & = & 0,
\label{eq:MHD_formalism:MHDdel_uB_zero}
\end{eqnarray}
where ${\bf u, b}, p$ are respectively the velocity, magnetic, and normalized pressure fields, $\Ru$ and $\Rm$ are kinetic and magnetic Reynolds numbers. $ {\bf F}_u$ and $ {\bf F}_b$ are the external large-scale forces.

The equations are numerically integrated using pseudospectral code \textsc{Tarang} \cite{Verma2013Tarang}. It is a general-purpose flow solver for turbulence and instability studies. \textsc{Tarang} scales nearly up to 196608 cores \cite{CHATTERJEE201877}. Helical
magnetohydrodynamic turbulence can be studied using implemented features: forcing, spectra, fluxes and mode-to-mode transfer of kinetic, magnetic and cross helicity \cite{Stepanov2017ISPRAS}.
For the purpose of this work we introduce  random forces which control the total energy injection $\varepsilon$ and cross helicity $\varepsilon_c$ rates. In Fourier space their expressions can be written in the following form
\begin{eqnarray}
\label{eq:forces:definition}
\bF_u(\bk) & = & ( (\varepsilon - \varepsilon_c)^{1/2}\be_u(\bk) + \varepsilon_c^{1/2} \be_c(\bk)), \\
\bF_b(\bk) & = & ( (\varepsilon - \varepsilon_c)^{1/2} \be_b(\bk) \pm \varepsilon_c^{1/2} \be_c(\bk)), \label{eq:forces:definition1}
\end{eqnarray}
where $\be_{*}$ -- random vector with $|\be_{*}|=(\Delta t N_f)^{-1/2}$, where `$\ast$' stands subscripts $u, b$ and $c$. Here $\Delta t$ is a time step, $N_f$ is a number of modes where forcing is applied. More general form of Eqs.~(\ref{eq:forces:definition}) and (\ref{eq:forces:definition1}), and their detailed derivations can be found in \cite{Titov2018CMM}. In our forcing parametrization $\varepsilon\geq\varepsilon_c\geq0$. Sign of injected cross helicity is controlled by the choice of sign in Eq.~(\ref{eq:forces:definition1}).

\section{Results.}

\begin{figure}
  \centering
  \hspace{1.0cm}{\bf\small (a)}\hspace{5.5cm}{\bf\small (b)}\\
  \includegraphics[width=0.49\columnwidth]{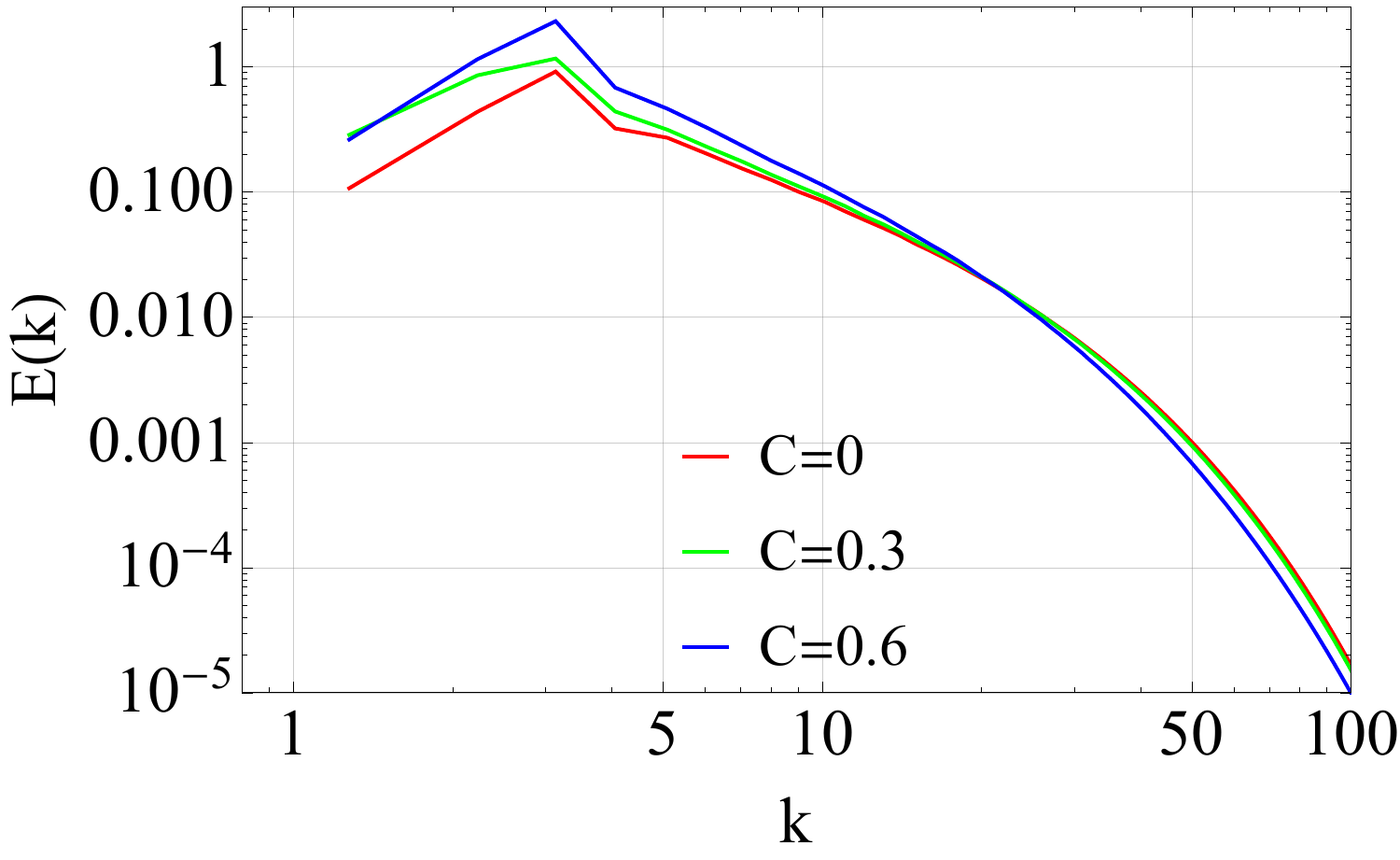}
  \includegraphics[width=0.49\columnwidth]{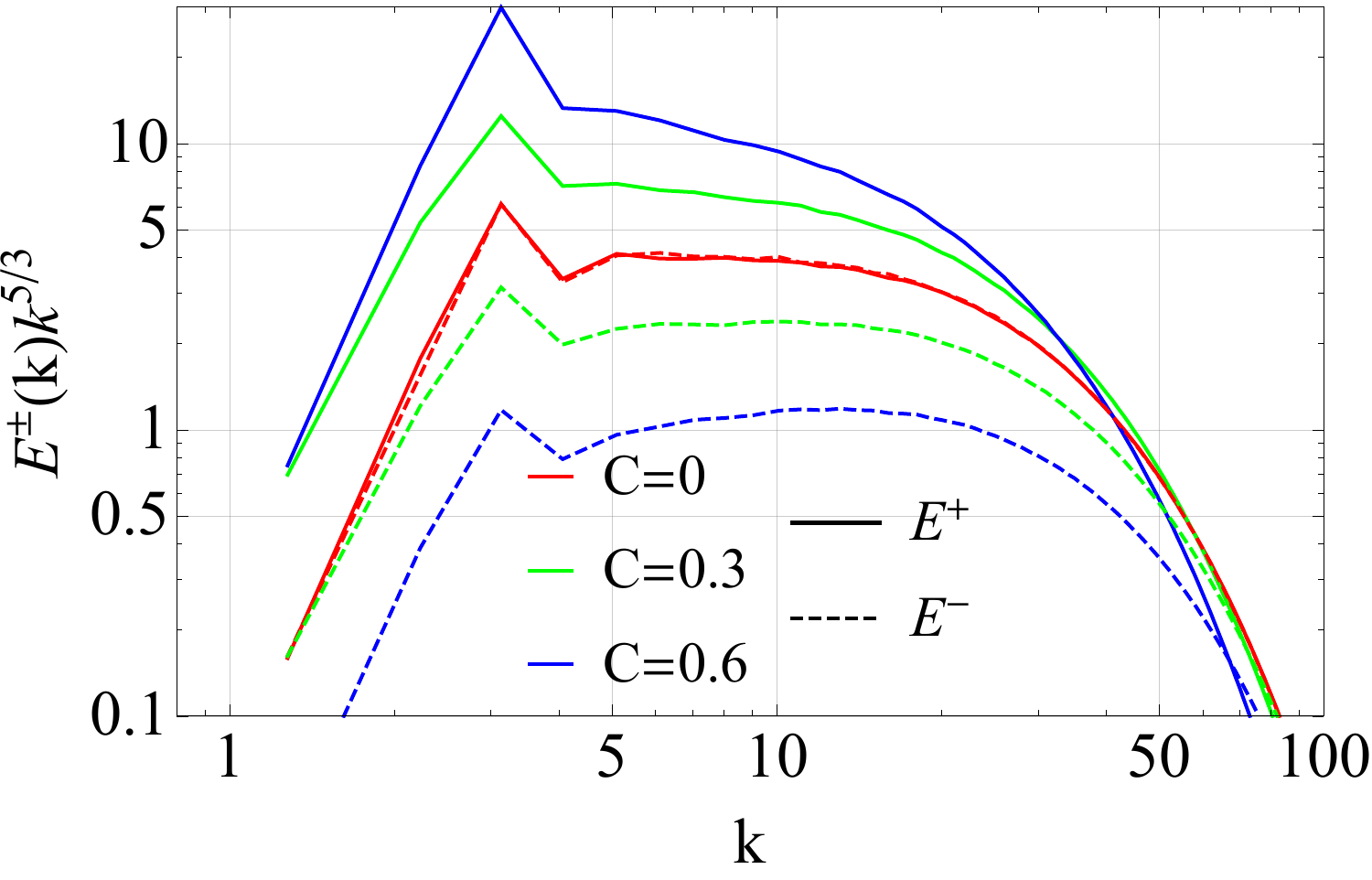}
  \caption{For MHD simulations with different cross helicity forcing ($C$=0, 0.3, 0.6): (a) Total energy spectra;  (b) the compensated energy spectra of Els\"{a}sser variables.  }
\label{fig1}
\end{figure}

\begin{figure}
  \centering
  \hspace{1.0cm}{\bf\small (a)}\hspace{5.5cm}{\bf\small (b)}\\
  \includegraphics[width=0.49\columnwidth]{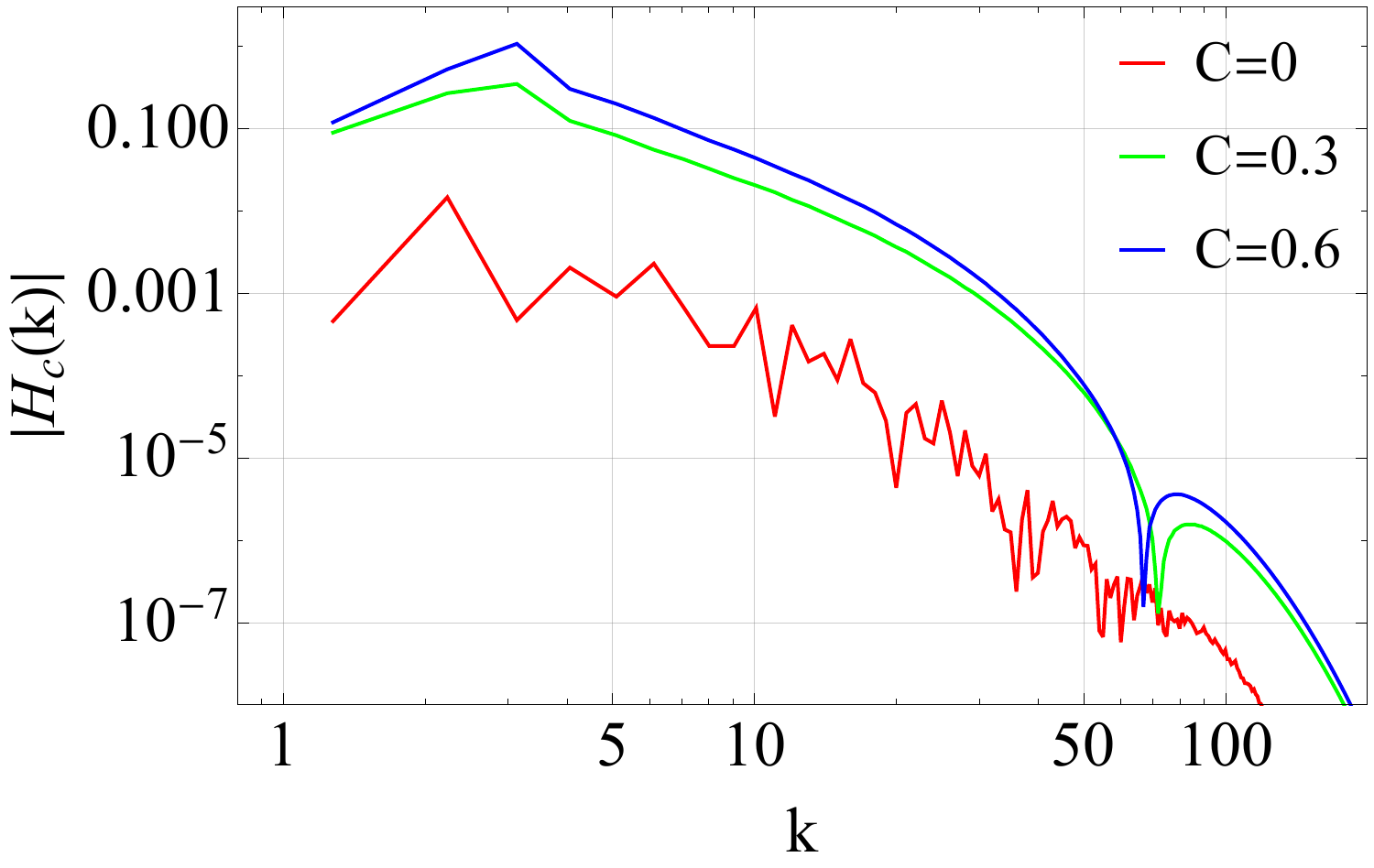}
  \includegraphics[width=0.49\columnwidth]{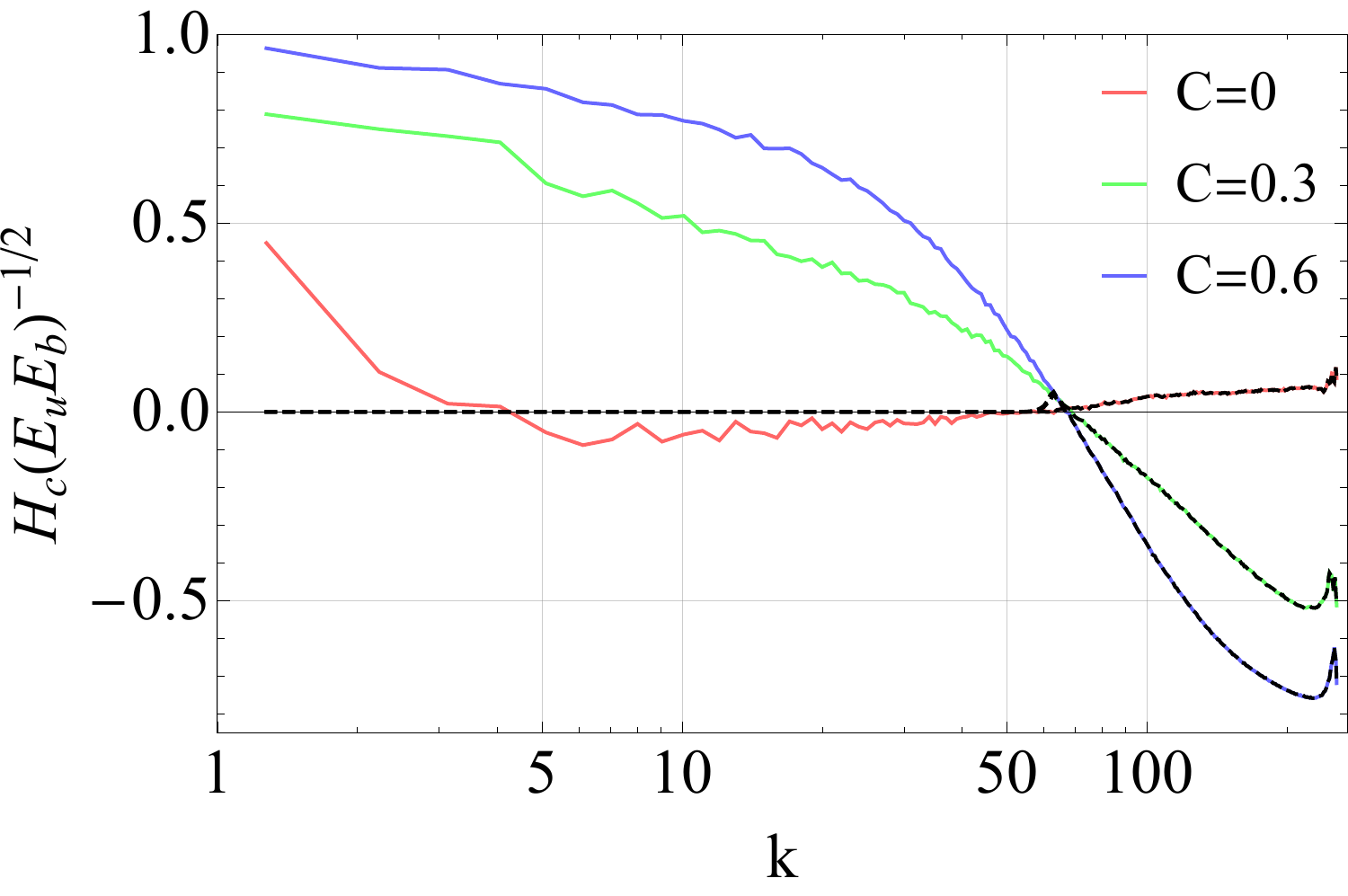}
  \caption{For MHD simulations with different cross helicity forcing ($C$=0, 0.3, 0.6): (a) Absolute value of cross helicity spectra; (b) compensated relative spectra of cross helicity. Dashed lines represent remnants after high pass filtering. }
\label{fig2}
\end{figure}

Simulations were performed in a triply periodic domain of size $2\pi$ on a grid $512^3$. Forcing acts in a range of scales $1 < |\bk| \le3$. We fixed $\varepsilon = 2$  with what we achieved  $\rm Re = Rm \approx 2094$. It corresponds Kolmogorov's dissipation wave number $k_d\approx100$. Parameter $C=\varepsilon_c/\varepsilon$ was varied to distinguish the effect of cross helicity. The simulation is initialized with random distribution of weak kinetic and magnetic fields. It takes about 20 units of time to reach a quasi-stationary regime of MHD turbulence. The data of next 10 units of time  are taken for evaluating the averaged energy and cross helicity spectra for three values of $C$=0, 0.3, 0.6 (see  Figs.~\ref{fig1} and \ref{fig2}).

Total energy spectra in Fig.~\ref{fig1}(a) get steeper with the increase of $C$. The energy is gradually accumulated at large scales because its transfer in the inertial range is suppressed by cross helicity. One can see that spectra cross each other at scale $k\approx20$. This meant that dissipation scales is larger for larger $C$. Such behavior has also been observed in MHD turbulence simulated with the shell models \cite{Mizeva:DP2009}.
Fig.~\ref{fig1}(b) shows separated energy spectra for the Els\"{a}sser variables $\bZ^{\pm}=\bu\pm\bb$. We find that $E^+$ increases but $E^-$ decreases compared to the $C=0$ case. It is remarkable that $E^-$ has a peak at the beginning of the dissipation scales. It can be explained by a fast drop of $E^+$ which works as advection for $E^-$.

Cross helicity also increases with $C$ but the slope  does not change significantly (see Fig.~\ref{fig2}(a)). One can observe an abrupt fall of $|H_c(k)|$ near $k_r\approx70$. It corresponds to a change of sign which is better visible in distribution of relative cross helicity $H_c^r=H_c/(E_u E_b)^{1/2}$  (Fig.~\ref{fig2}(b)). We note this remarkable feature of $H_c$ that it changes its sign and reaches rather high relative values at dissipative scales. Sign reversal by $H_c$ reflects a crossing of $E^+$ and $E^-$ since $H_c=E^+-E^-$. It may happen due to a tendency of $E^+$ to steepen and $E^-$ to flatten with increase of $C$.

\begin{figure}
  \centering
{\bf \small (a)}\hspace{3.7cm}{\bf\small (b)}\hspace{3.7cm}{\bf\small (c)}\\
  \includegraphics[width=0.32\columnwidth]{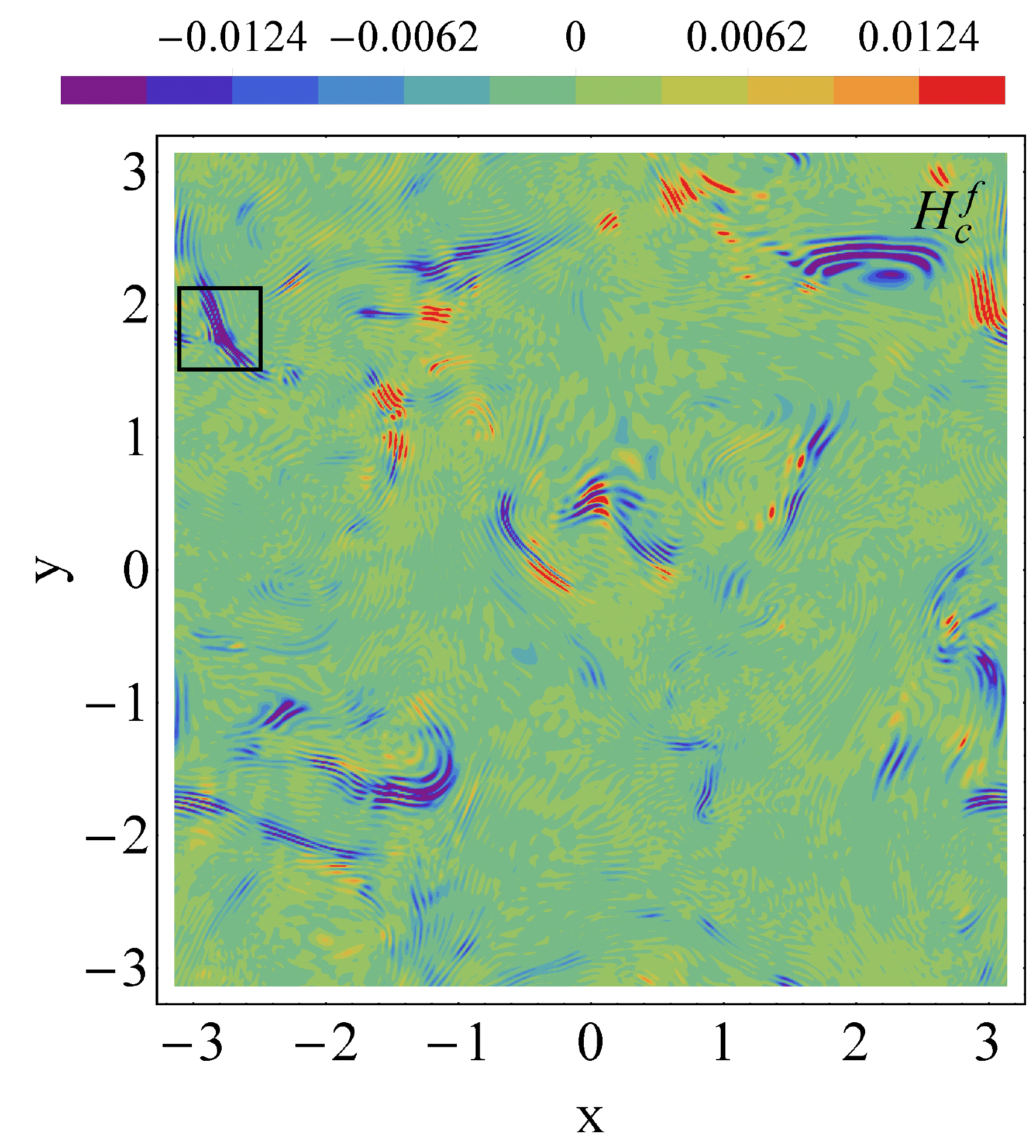}
  \includegraphics[width=0.32\columnwidth]{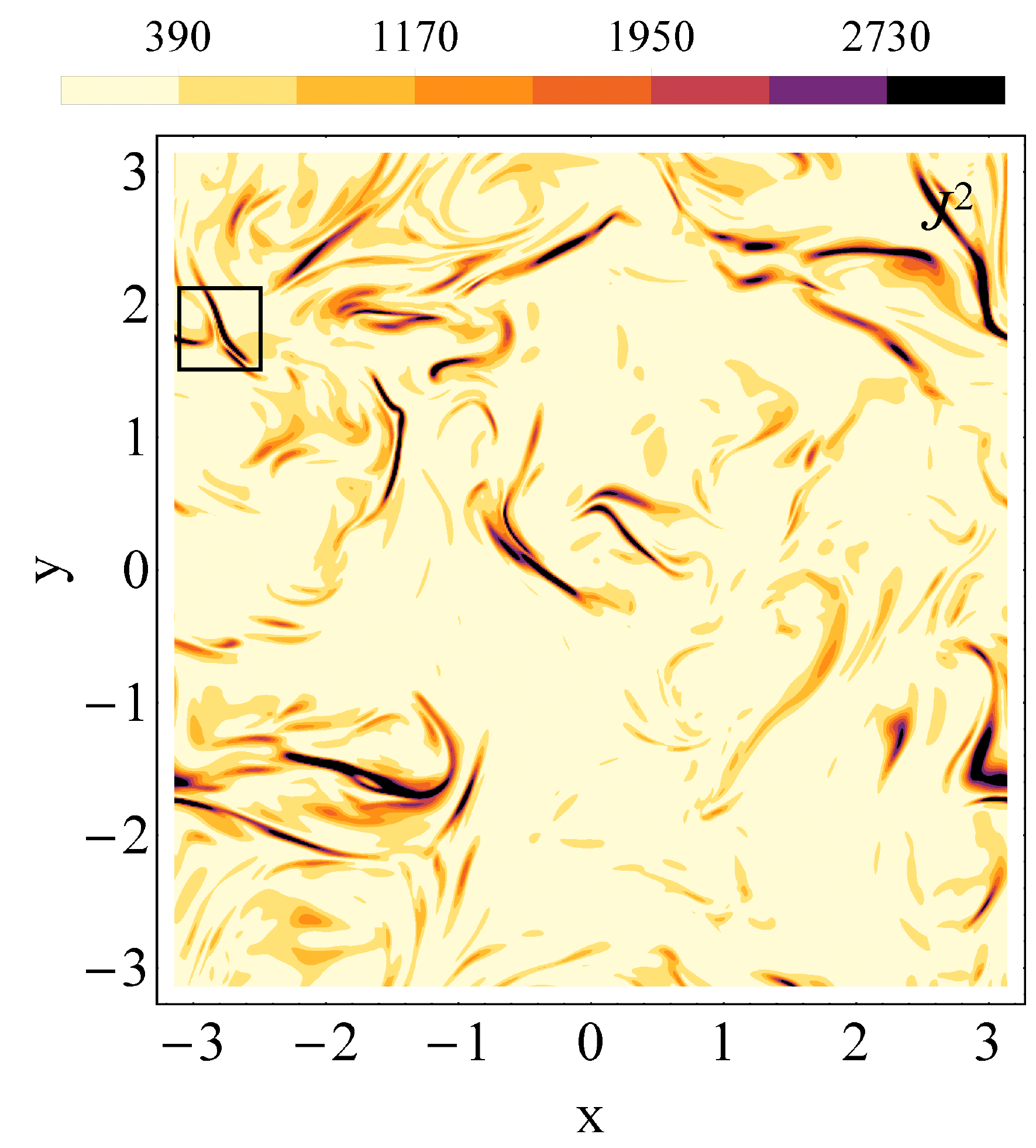}
  \includegraphics[width=0.32\columnwidth]{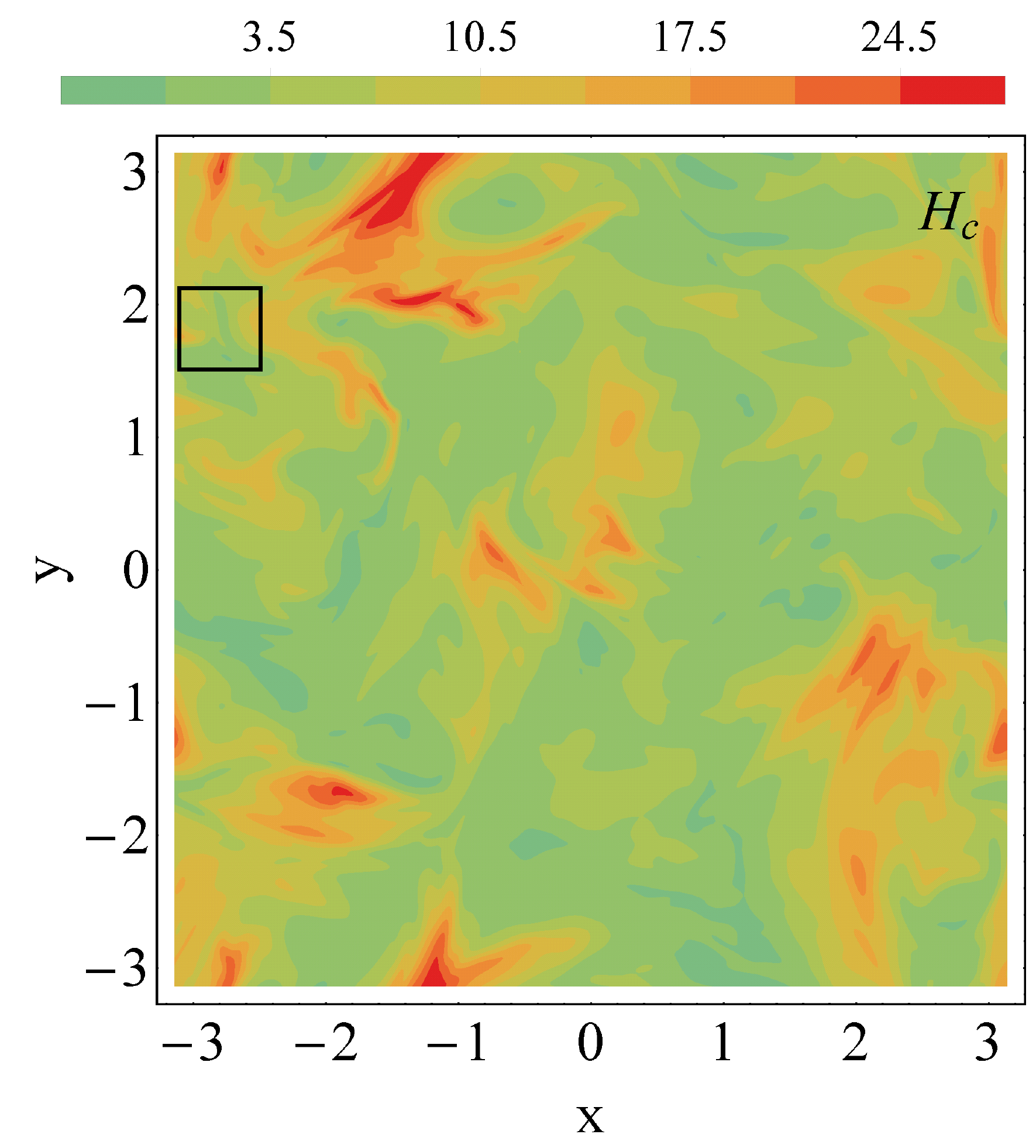}
  \caption{For MHD simulation with $C=0.6$, density plots at a horizontal cross section at z=1.78 of : (a) Small scales of cross helicity; (b) dissipation rate of total magnetic field; (c) total cross helicity.  The distributions are shown in a range limited by $\pm \frac{1}{10}$ of maximum absolute value of $H_c^f$ for panel (a) and by $\pm \frac{1}{2}$ of maximum value of $|J^2|$ for panel (b).}
\label{fig3}
\end{figure}

We extract small scale kinetic and magnetic fields by applying a Gaussian filter in Fourier space as
\begin{equation}\label{filter}
  \bx^f(\bk)=\bx(\bk){\rm e}^{-(k_r-k)/k_f}
\end{equation}
for all $k<k_r$ with a filter width $k_f=3$.
Instantaneous spacial cross-section distribution of small-scale cross helicity $H^f_c=\bu^f\cdot\bb^f$ is shown in Fig.~\ref{fig3}(a), and that of the magnetic dissipation $|{\bf J}|^2$ is shown in Fig.~\ref{fig3}(b).
In Fig. 3(a) we observe strong peaks of negative cross helicity (opposite to the sign of the injected $H_c$).
$H_c$ is localized in the filamentary structures. The regions of negative $H_c$ is  strongly correlated with those with strong $|\bJ|^2$ (see Fig.~\ref{fig3}(b)), which describes a magnetic dissipation and reconnection. Fig.~\ref{fig3}(c) shows $H_c$ distribution of all scales that are dominated by large-scale forcing.

\begin{figure}
  \centering
  \includegraphics[width=0.32\columnwidth]{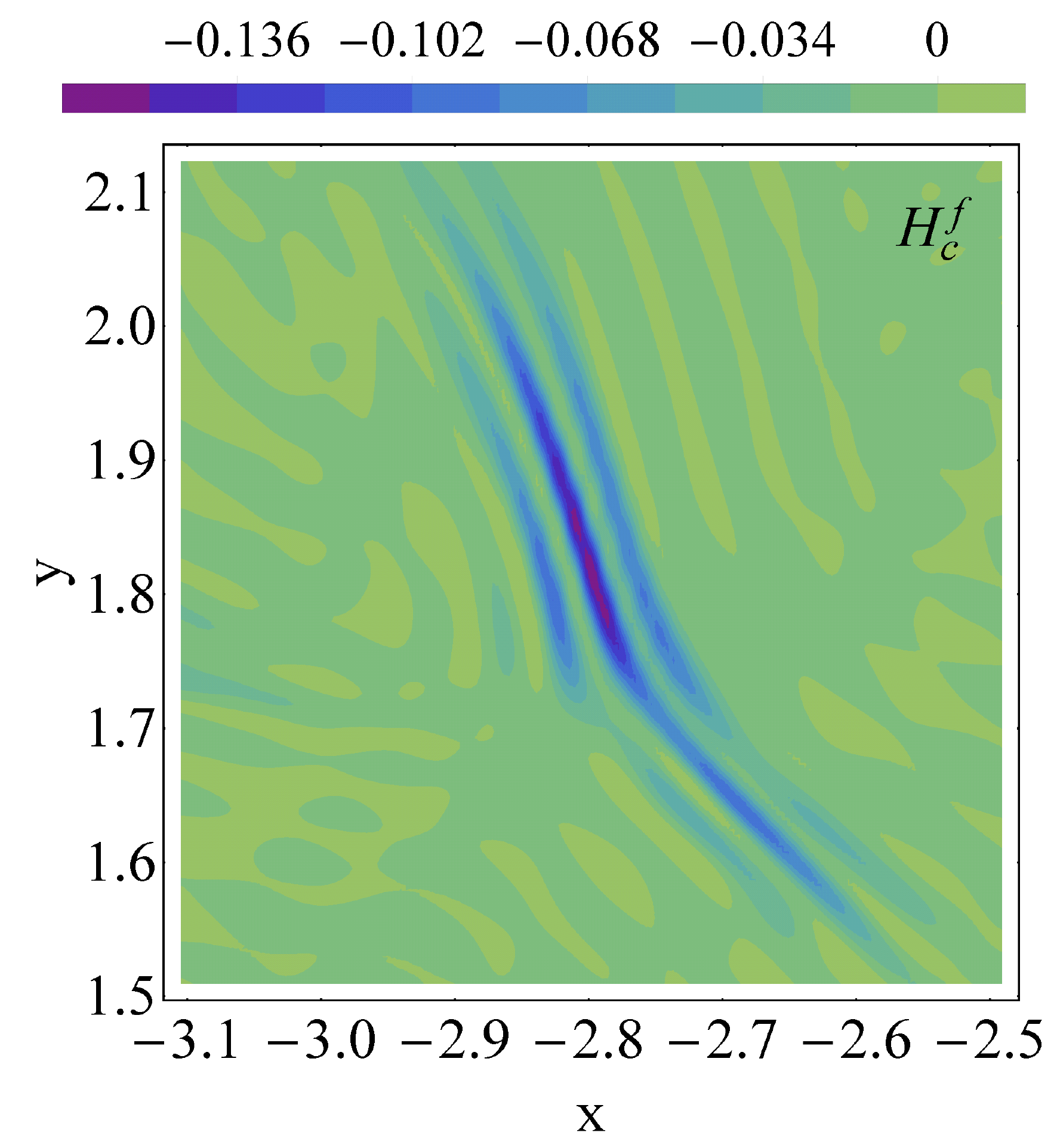}
  \includegraphics[width=0.32\columnwidth]{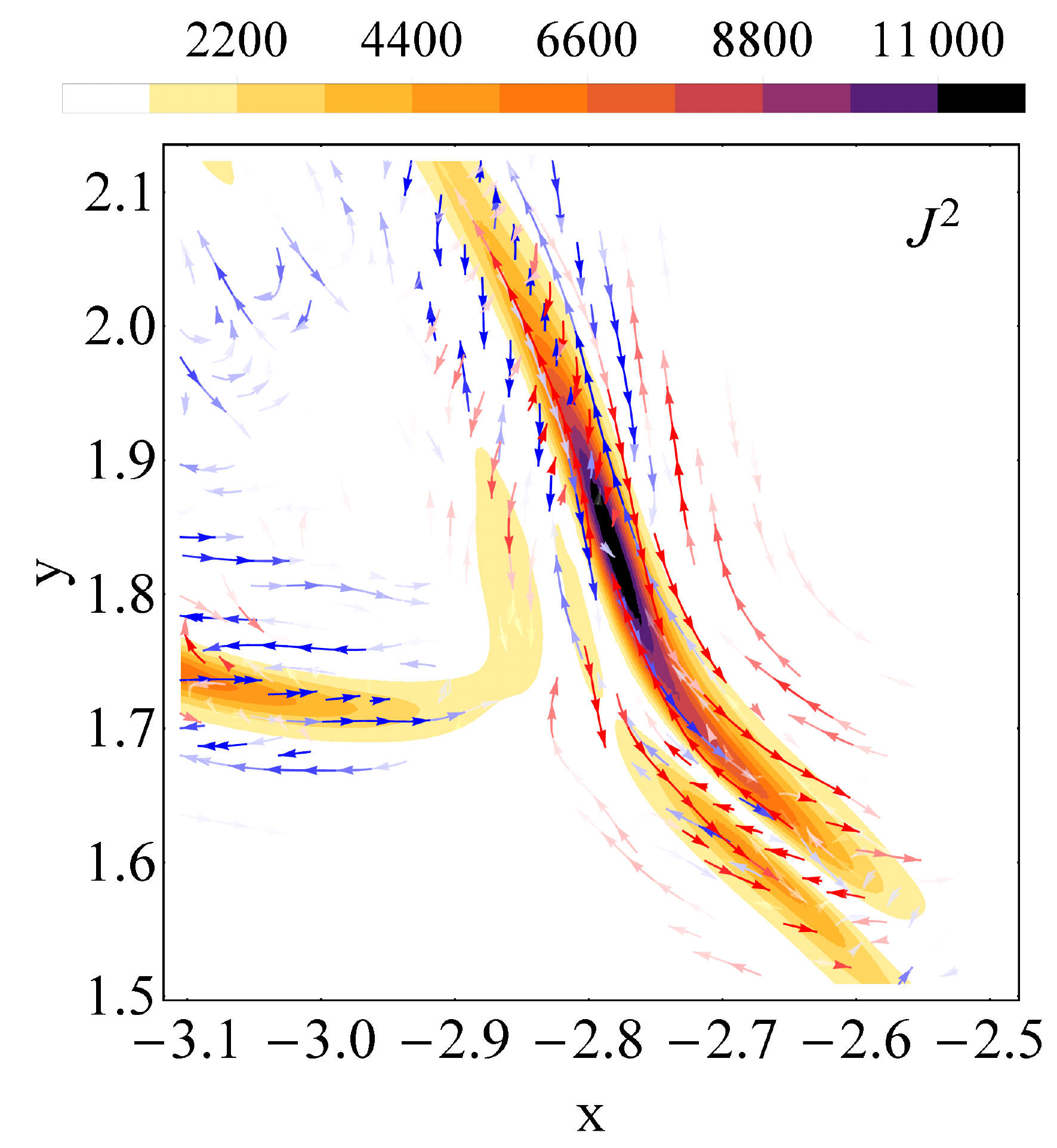}
  \includegraphics[width=0.32\columnwidth]{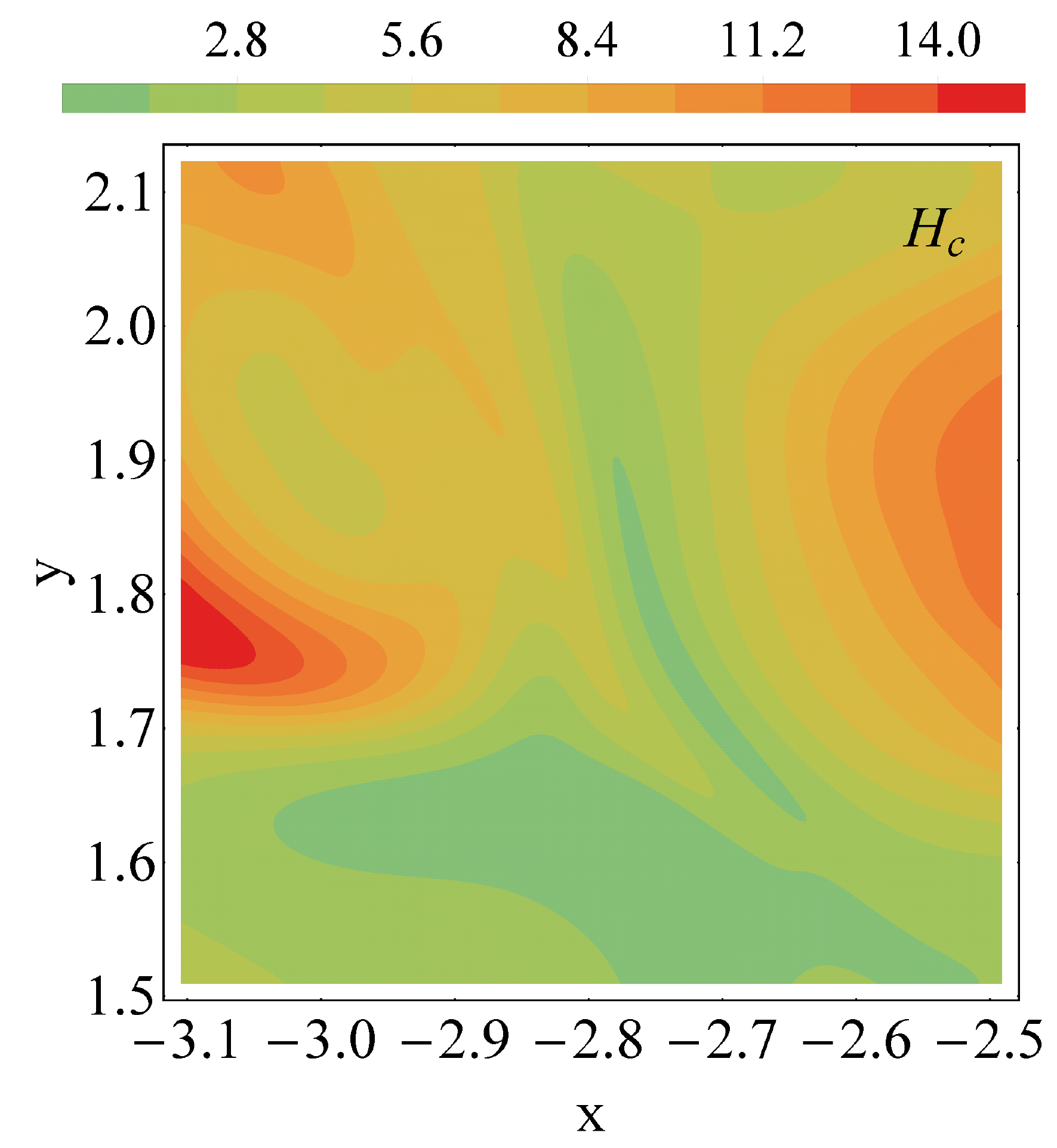} \\
  \includegraphics[width=0.32\columnwidth]{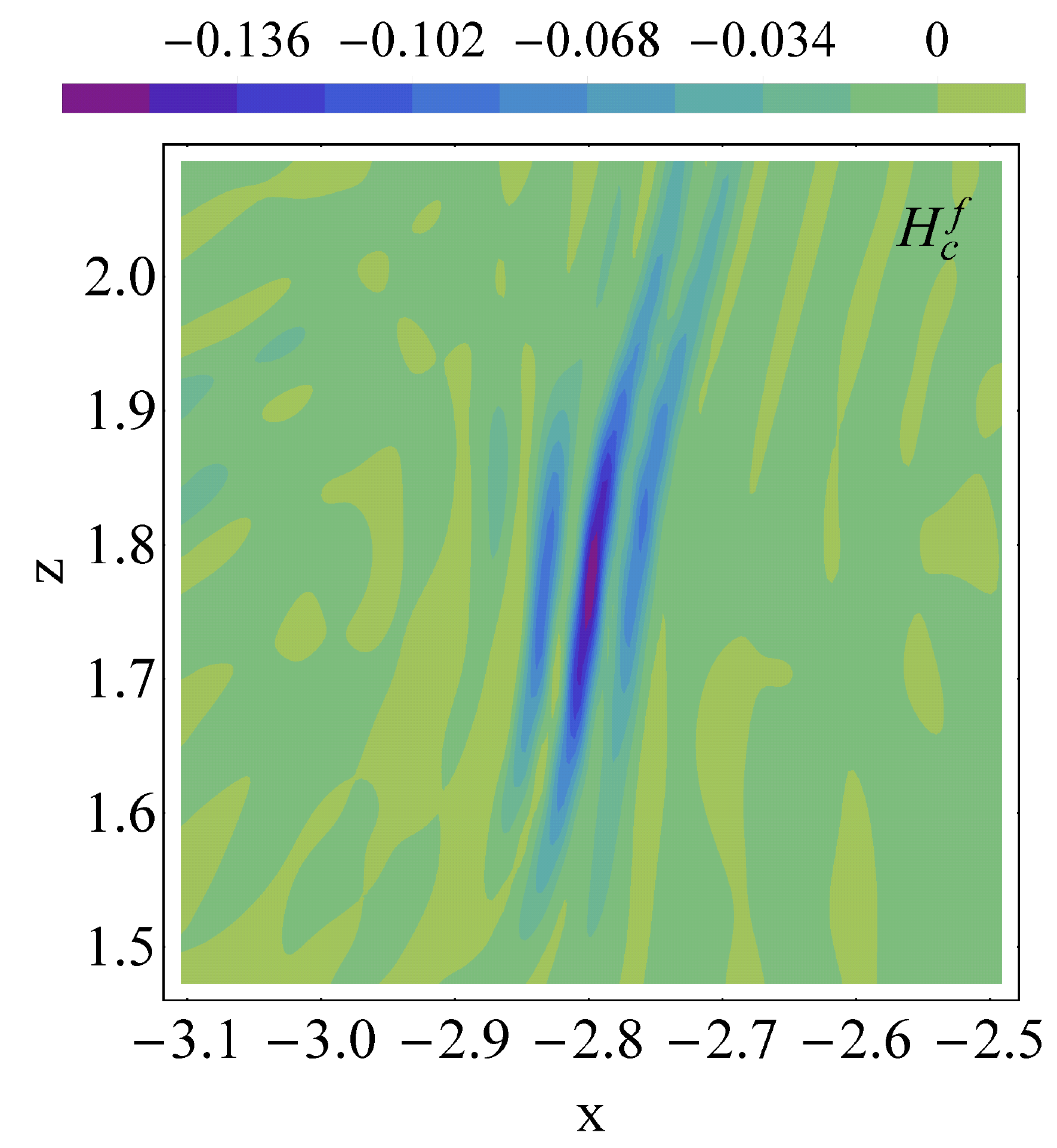}
  \includegraphics[width=0.32\columnwidth]{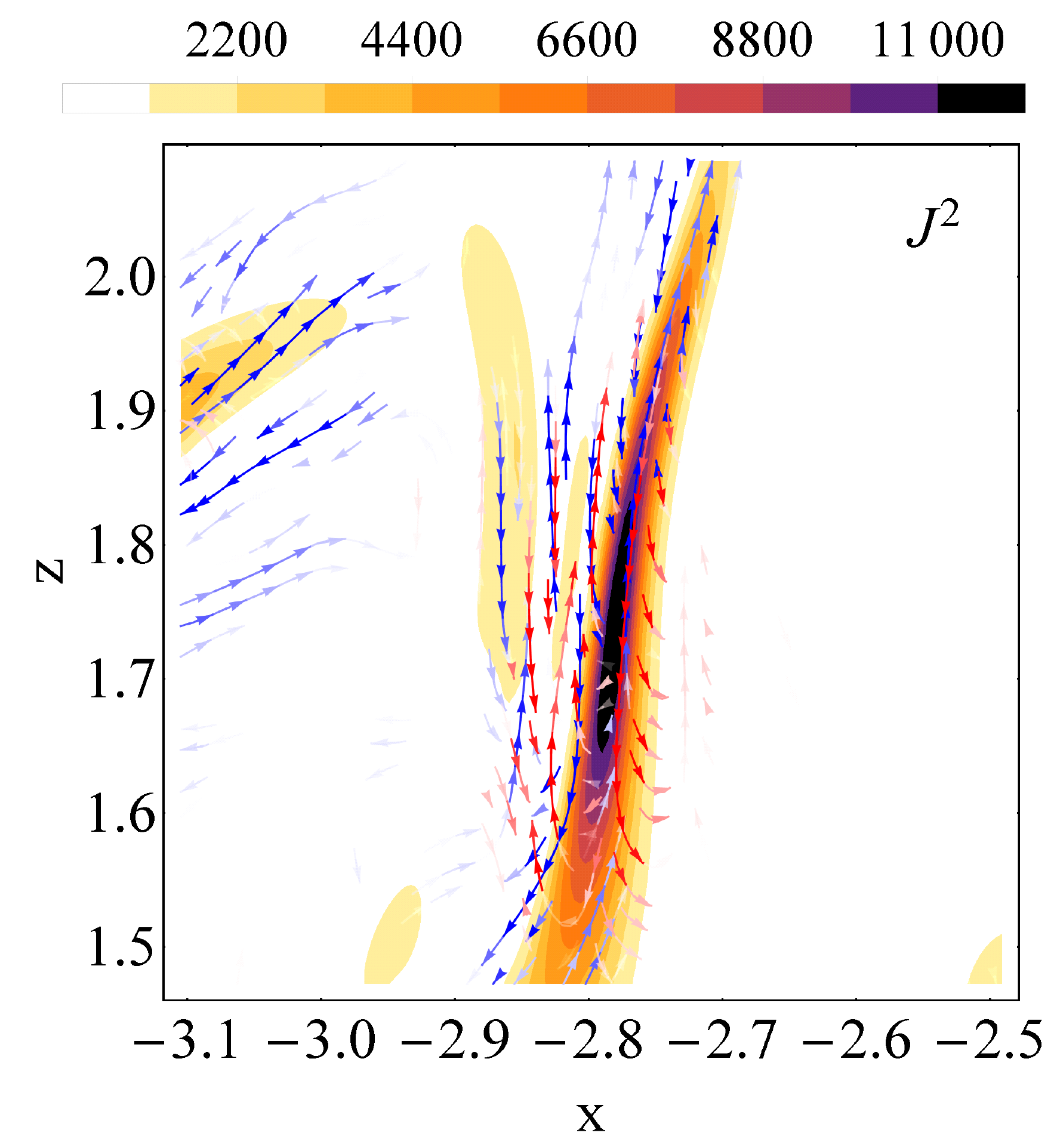}
  \includegraphics[width=0.32\columnwidth]{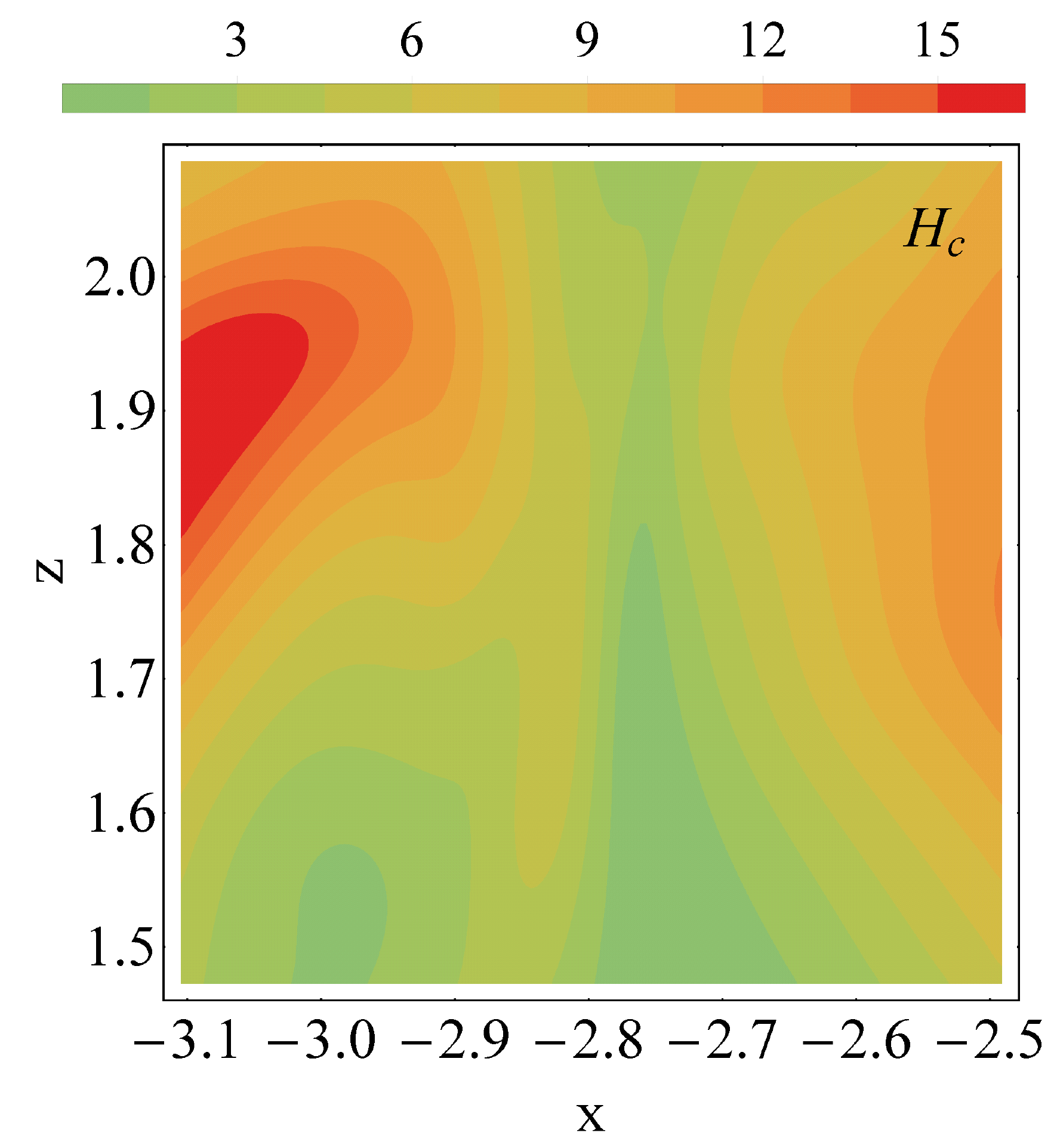} \\
  \includegraphics[width=0.32\columnwidth]{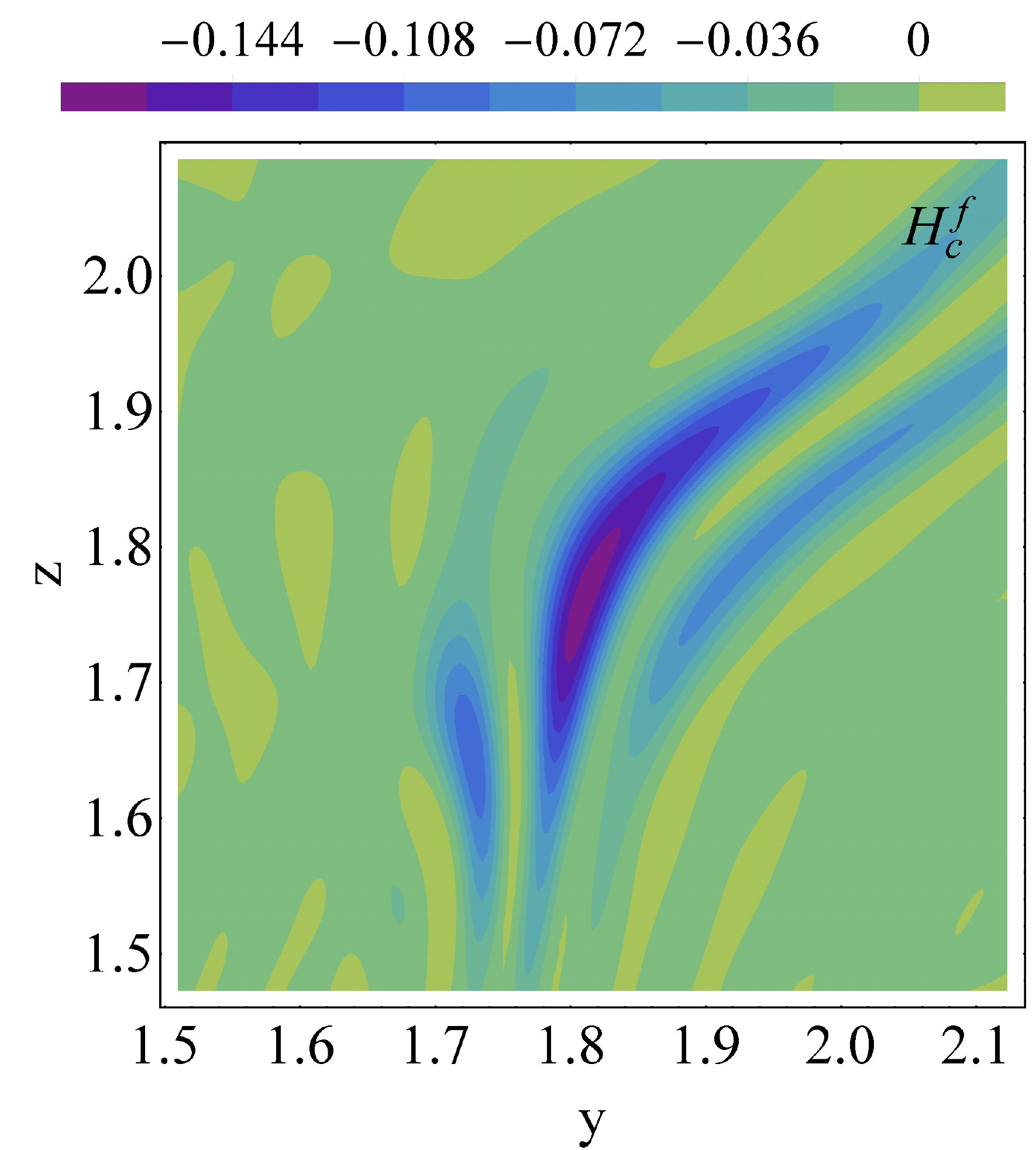}
  \includegraphics[width=0.32\columnwidth]{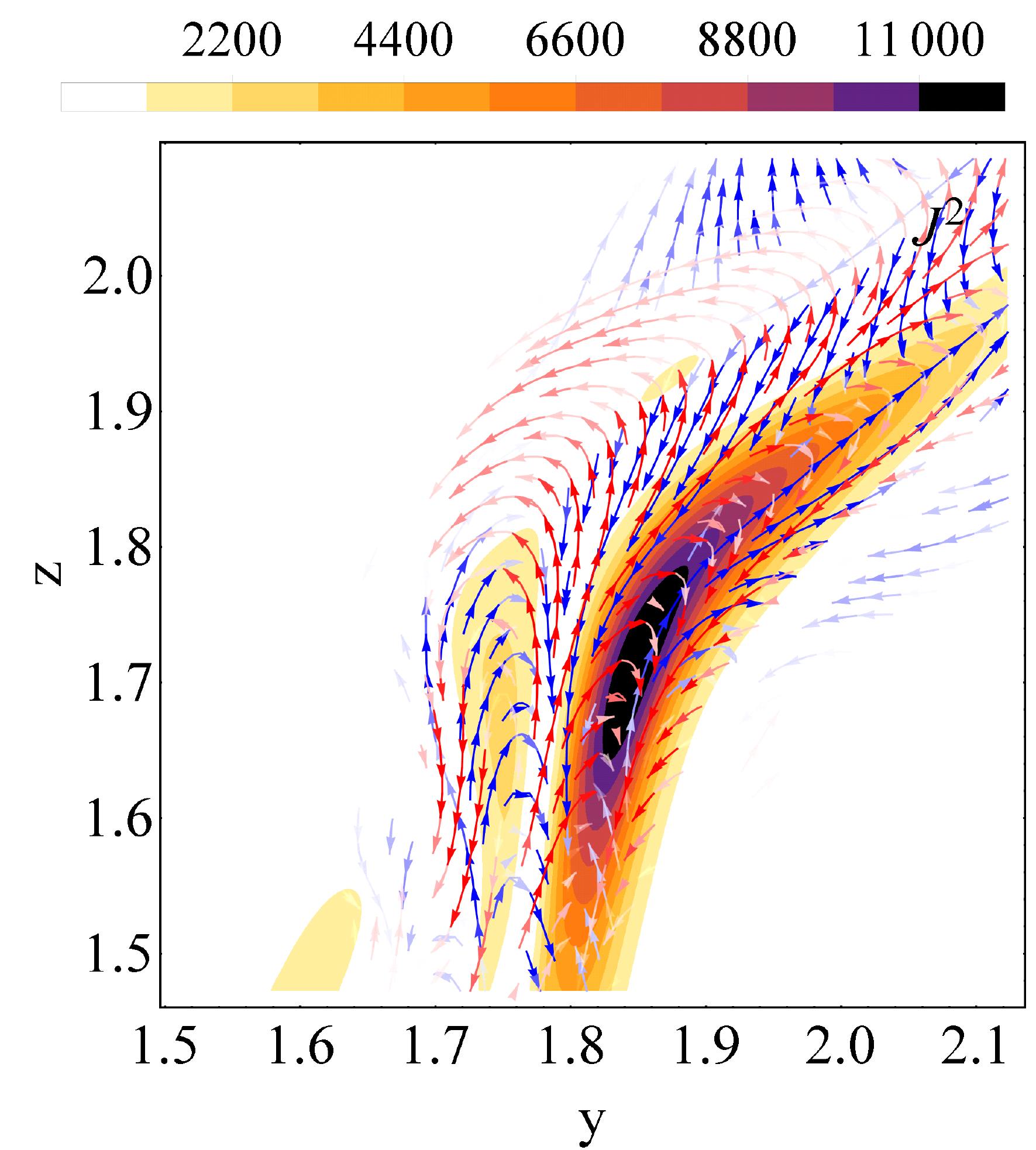}
  \includegraphics[width=0.32\columnwidth]{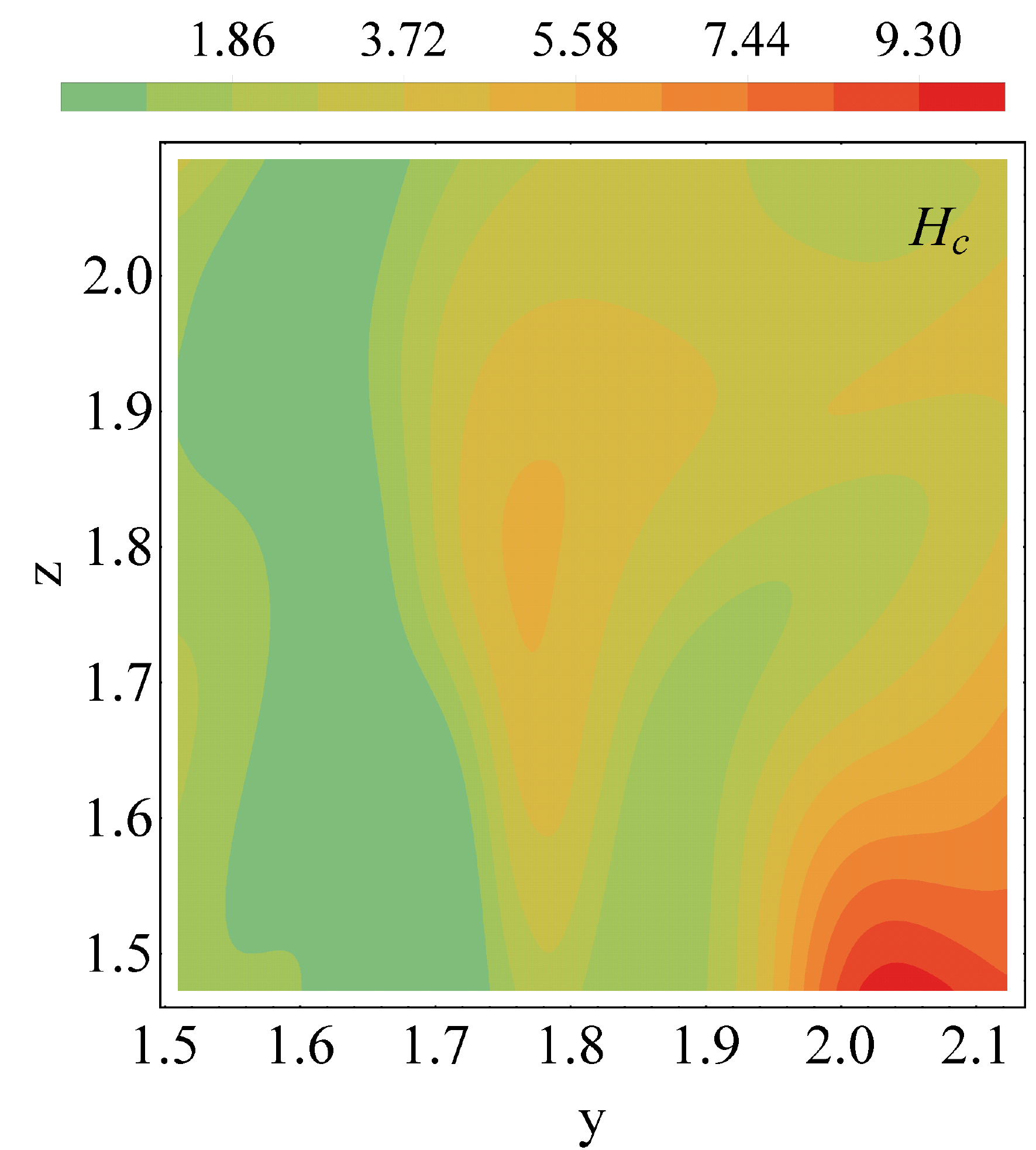}
  \caption{For MHD simulation with $C=0.6$, at the cross-sections  $xOy$, $yOz$ and $xOz$ (from top to bottom): (left column) density plots of $H_c^f$; (middle column) streamlines of small scale magnetic field (red) and velocity field (blue), and density plots of magnetic dissipation $|\bJ|^2$; (right column) density plots of cross helicity $H_c$  in the vicinity of the selected reconnection zone enclosed by rectangular boxes in Fig.~\ref{fig3}.}
\label{fig4}
\end{figure}

We consider in details an anisotropic structure of an individual reconnection. In Fig.~\ref{fig4} we exhibit the distribution of small scale kinetic and magnetic fields, total $H_c$ and $|\bJ|^2$ in orthogonal cross-sections  enclosed by rectangular boxes (upper left regions) in Fig.~\ref{fig3}.


\section{Conclusions}

We numerically investigated the distribution of cross helicity in MHD turbulence that was forced with  kinetic energy and cross helicity injections at large scales. In the dissipation range, we observe a significant level of the relative cross helicity, as well as reversal of cross helicity as we traverse from large scales to small scales.  This feature has been reported earlier in ~\cite{PhysRevFluids.2.114604}, but it was not analysed in detail.   The presence of relatively large  levels of relative cross helicity in the dissipation range is in marked contrast to the vanishing kinetic helicity in dissipative range of hydrodynamic turbulence.  This difference is because of the difference in the order of derivatives of kinetic helicity and cross helicity in the dissipation range.

More importantly, we observe that magnetic reconnections occur at the regions where small scale negative cross appears.  Also, the length scales of the cross helicity  and the magnetic reconnection are very similar.  If we changes the scales (or corresponding modes in the shell model), this correspondence between the vanishing cross helicity and the high dissipation becomes less clear.  This issue merits further investigation.

At a scale other than the cross-helicity reversal, the cross-helicity density in turbulence is in general finite and not null. The scale dependence of the spatiotemporal distributions of the cross helicity and magnetic dissipation at very high Reynolds numbers can be effectively done only with the shell models of MHD turbulence~\cite{Stepanov2013,2010EL.....9234007F}.

\section*{Acknowledgment}
We are grateful to Professor Franck Plunian for his interest in the subject and useful comments. The mathematical model development and numerical simulations performance was supported  by the  Department of Science and Technology, India (INT/RUS/RSF/P-03) and Russian Science Foundation (RSF-16-41-02012) for the Indo-Russian project.  Computational resources were provided Shaheen II of the Supercomputing Laboratory at King Abdullah University of Science and Technology (KAUST) under the project K1052. Theoretical studies magnetic reconnection problems was also supported by the Japan Society for the Promotion of Science (JSPS) Grants-in-Aid for Scientific Research 18H01212. We thank the scientific committee of the Third Russian Conference on Magnetohydrodynamics for organising a warm atmosphere and stimulation discussions.

\bibliographystyle{mhd}
\bibliography{ref}

\lastpageno
\end{document}